\documentstyle[11pt,newpasp,twoside,epsf]{article}
\newcommand{\hmp}{h^{-1}Mpc}  
\def\spose#1{\hbox to 0pt{#1\hss}}   
\def\ltapprox{\mathrel{\spose{\lower 3pt\hbox{$\mathchar"218$}}   
 \raise 2.0pt\hbox{$\mathchar"13C$}}}   
\def\gtapprox{\mathrel{\spose{\lower 3pt\hbox{$\mathchar"218$}}   
 \raise 2.0pt\hbox{$\mathchar"13E$}}}   
\def\inapprox{\mathrel{\spose{\lower 3pt\hbox{$\mathchar"218$}}   
 \raise 2.0pt\hbox{$\mathchar"232$}}}

\def\edcomment#1{\iffalse\marginpar{\raggedright\sl#1\/}\else\relax\fi}
\reversemarginpar

\begin{document}
\title{Scaling and fluctuations in galaxy distribution:
two tests to probe large scale structure}
 \author{Francesco Sylos Labini}
\affil{ D\'ept.~de Physique Th\'eorique,
	 Universit\'e de Gen\`eve, Quai E. Ansermet 24, 
CH-1211 Gen\`eve, Switzerland 
and INFM Sezione Roma1, Roma  Italy}
\author{Andrea Gabrielli}
\affil{Lab. PMC, Ecole Polytechnique, 91128 - Palaiseau Cedex,
 France}

\begin{abstract}
We present a brief introduction to the statistical
properties of systems with large fluctuations.
We point out that for such systems the relevant
statistical quantities are scaling exponents and the nature
of fluctuations is completely different from the 
one belonging to system with small fluctuations.
We then propose two test to be performed
on galaxy counts data as a function of 
apparent magnitude, which may clarify in a 
direct way the nature of the large scale
galaxy clustering
\end{abstract}

\keywords{cosmology, galaxy distribution, statistical methods}

In Fig.1 it is shown an homogenous distribution
with poissonian fluctuations in the two-dimensional
Euclidean space. In such a distribution 
the average density is a well defined property
if it measured on scales larger than the mean particle
separation $\Lambda$.
\begin{figure}  
\vspace{4cm}
\plotfiddle{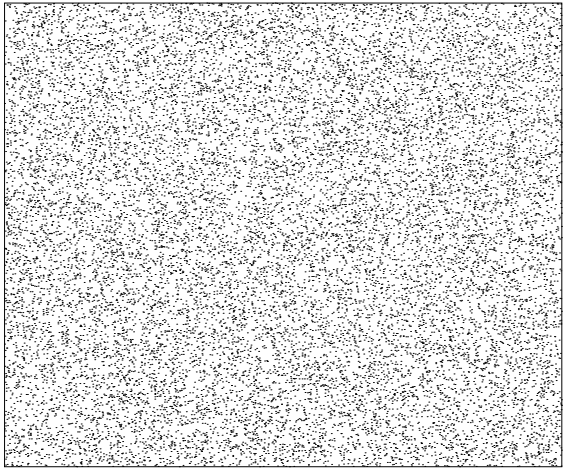}{4cm}{0}{50}{50}{-180}{-30}
\caption{ Homogeneous distribution 
with Poisson fluctuations}
\end{figure}
Even if there are some {\it small amplitude fluctuations} up to a certain
scale, or up to the sample's size, the average density is still
a well defined property. For example in Fig.2 it is shown 
a distribution where the fluctuation structures are extended
over the whole sample. In such a case the average density is
still a well defined property, and it can be measured
at large enough scales ($r > \lambda_0$), 
when the {\it amplitude}
of the fluctuations become smaller than the average
density itself, i.e. when 
$\delta N(r) /\langle N(r) \rangle \le 1$.
\begin{figure}  
\vspace{4cm}
\plotfiddle{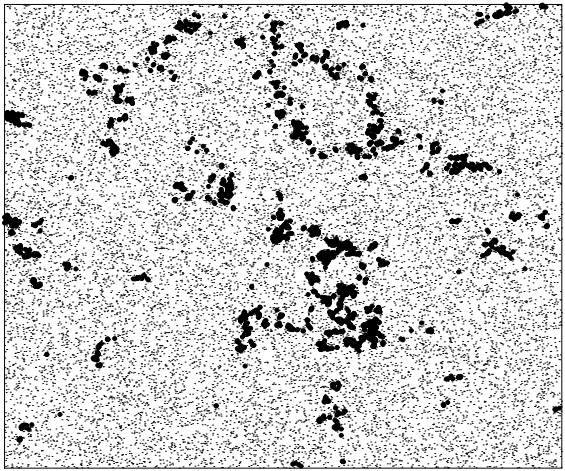}{4cm}{0}{50}{50}{-180}{-30}
\caption{Homogeneous distribution with small amplitude
long range correlations}
\end{figure}
These two systems belong to the class of distributions with small
amplitude fluctuations. The fact that the 
average density is well defined
implies  that any  one point property is well defined.
For example in a
ball of radius $r>\lambda_0>\Lambda$ centered in a {\it randomly } 
chosen point, in both 
the distributions, the number of points is constant, 
a part a small fluctuations whci vanishes
when $r$ is farther increased.  In these case when the 
average density is well-defined,  it
is possible to study the statistical properties of the fluctuations
around the it.
For example, given an occupied point, the number of
points at distance $r$ is given by (Peebles, 1980)
\begin{equation}
\label{e1}
N_p(<r) = \langle n \rangle V(r) + \langle n \rangle \int_V \xi(r) d^3r
\end{equation}
where 
\begin{equation}
\label{e2}
\xi(r) \equiv \frac{\langle n(r) n(0)\rangle}{\langle n \rangle^2} -1
\end{equation}
Note that the volume integral of $\xi(r)$, over the whole space
where the average density $\langle n \rangle$ (one point property) 
has been estimated,
must be zero by definition. In a finite volume,
the integral of $\xi(r)$ gives the excess of points
with respect to a Poisson distribution.

A completely different situation happens in the system 
as the one shown in Fig.3.
\begin{figure}  
\vspace{4cm}
\plotfiddle{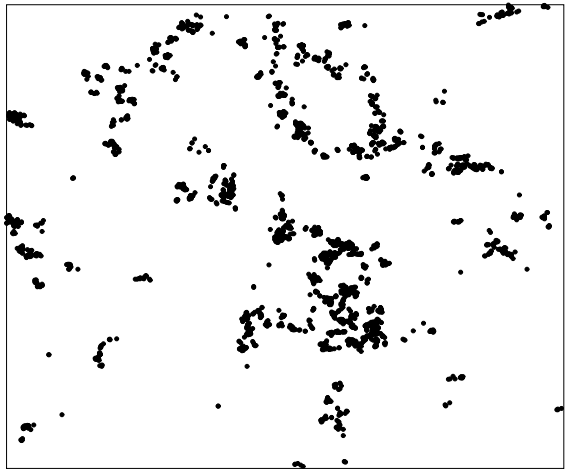}{4cm}{0}{50}{50}{-180}{-30}
%\plotone{sylos3.eps} 
\caption{Fractal distribution with $D=1.47$ in the two
dimensional Euclidean space}
\end{figure}
Suc a 
system is characterized by having voids of 
all sizes up to the sample size.
For such a distribution, it is not possible
to define the average density, or more generally, to study
the one point properties. This is because the 
system is characterized by having structures and voids of 
all sizes and at all scales (up to the sample size).
Hence if we make the simple exercise we did before, 
by putting a ball of radius $r$ centered in a randomly chosen
point, the number of points inside the ball
will strongly fluctuate from a position to another.
For example, if the center is in a void of size
larger than the ball's diameter, we do not 
find any point, while if the center is in a 
large scale structure we find a huge number of points.
This happens as long as the distribution has fractal
properties, i.e. as long as the size of the voids
increases in a proportionally to the size
of the sample. Hence in such a situation 
one will never find a scale $\lambda_0$ above which
the fluctuations become small and it is not 
possible
to measure the one-point average density.
Fractals are then distribution with  fluctuations
as large as the (conditional) average density at all scales.
The proper way to characterize the statistical 
properties is to focus on the scaling behavior
of two-point quantities. For example 
one may study the two-point conditional
average density (Pietronero 1987, Sylos Labini et al., 1998) defined as
\begin{equation}
\label{e3}
\Gamma(r) = \frac{\langle n(r) n(0) \rangle}{\langle n \rangle} \sim r^{D-3}
\end{equation}
This function measures the average density at distance $r$ 
from an {\it occupied point}. 
The term $\langle n \rangle$ is just a normalizing
factor equal to $N/V$ which is useful in order to compare
the results from different distributions 
where the number of points $N$ and the volume $V$
can be different. The last equality in Eq.3
gives the point of this discussion: if $D<3$ the 
conditional average density has a scaling 
behavior as a function of distance 
and hence: (i) the one-point average density is not 
a defined quantity and (ii) the only charactering
quantity is the scaling exponent or fractal
dimension $D$. If, otherwise, $D=3$
then the system has an average density defined,
and hence once can focus the attention to the 
statistical properties of the fluctuations
about the average density itself.
In the fractal case the distribution is 
characterized by {\it large fluctuations}.

These simple but fundamental properties 
of systems with large fluctuations change
our perspective on the studies of 
the statistical properties of a given distribution.
Instead of assuming that the one-point properties
are defined in any given sample, as it is done in the standard
analysis (Peebles 1980), we test   whether the 
distribution has such a property (Pietronero 1987, Sylos Labini et al., 1998).
The application of these methods to the redshift space
distribution of galaxies has been discussed at lengthly
and we refer to Sylos Labini et al. (1998) for a comprehensive
review of the subject.
Recently (Sylos Labini \& Gabrielli, 2000) we have proposed
to study  two properties of 
galaxy counts as a function of apparent magnitude,  
in order to discriminate   
between a small scale ($\sim 5 \div 20 \hmp$) homogeneous  
distribution, and  a fractal structure on large scales  
($\sim 100 \div 300 \hmp$):  
(i) The slope $\alpha$ of the counts  
can be associated, in the bright end where $z \ll 1$,
to a possible fractal dimension in real   
space, by simple arguments.  
(ii)  Fluctuations of counts  
about the average behavior as a function of apparent magnitude  
in the whole magnitude range $B \gtapprox 11^m$.  
These fluctuations are related to the very  
statistical properties  
of the spatial distribution, 
{\it independently on cosmological corrections}.  
While in an homogeneous distribution
they are exponentially damped as a function
of apparent magnitude, in a fractal 
they are persistent at all scales.
Large field-to-field fluctuations in the counts
have been reported both in the bright and in the faint
ends and in different photometric bands:
they can be due to some systematic measurement errors
or to a genuine effect of large scale structures. 
For this reson we need better calibrated photometric samples.
Therefore, the application of these tests to the forthcoming 
photometric surveys is then {\it fundamental}
to understand and characterize the 
large scale galaxy clustering.

%%%%%%%%%%%%%%%%%%%%%%%%%%%%%%%%%%%%%%%%%%%%%%%%%%%%%%%%%%%%%%%%%%%%%%%%%%%%%%%%%

\end{document}